\documentstyle[12pt,a4]{article}

\begin{document}
\date{\ }
\begin{center}
   {\large\bf The Hofstadter Energy Spectrum for an Interacting 2DEG.}\\
         {Vidar Gudmundsson$^1$ and Rolf R. Gerhardts$^2$\\
         $^1$Science Institute, University of Iceland, Dunhaga 3,\\
         IS-107 Reykjavik, Iceland. vidar@raunvis.hi.is\\
         $^2$Max-Planck-Institut f\"ur Festk\"orperforschung,
         Heisenbergstra{\ss}e 1,
         D-70569 Stuttgart,\\  Federal Republic of Germany.
         gerha@klizix.mpi-stuttgart.mpg.de}
\end{center}

\begin{quotation}
          We study the effects of the Coulomb interactions
          between electrons
          on the Hofstadter butterfly, which characterizes
          the subband structure of the Landau levels
          of a two-dimensional electron gas in a perpendicular
          homogeneous magnetic field and a periodic lateral
          superlattice potential.
          The interactions essentially preserve the intricate
          gap structure of the Hofstadter spectra,
          but with a lower symmetry that depends on the
          filling of the Landau bands. For short enough
          periods and strong enough modulation
          the miniband structure
          can be resolved in the thermodynamic density of states.
\end{quotation}

Measurements of the magnetotransport
in superlattices weakly modulated in one or two directions have
revealed novel oscillations in the conductivity, the Weiss oscillations,
reflecting the commensurability of the cyclotron radius
$R_c=l^2k_F=v_F/\omega_c$ of the electrons at the Fermi energy
$E_F=\hbar^2k^2_F/(2m^*)$ and the modulation
period $L$; $l=(c\hbar /eB)^{1/2}$ is the magnetic
length, and $\omega_c=eB/(mc)$ the cyclotron frequency
\cite{Gerhardts89:1173}.
The Weiss oscillations are superimposed on the Shubnikov-de Haas
oscillations which are well known in a homogeneous
two-dimensional electron gas (2DEG) and caused by the
commensurability of the Fermi wavelength $\lambda_F$ and $l$.
The single-particle energy spectrum for
this problem has been investigated by several
researchers
culminating in Hofstadter's butterfly, a graph
showing the complicated self-similar splitting
of a Landau band (LB) into minibands
as a function of the magnetic flux through a unit cell of the
lattice (for a review see \cite{Pfannkuche92:12606}).
The effects of the splitting of LB's into minibands has,
to the best of our knowledge, not been observed directly  in the currently
common superlattices ($L\sim 300\: $nm $n_s=3\times 10^{11}\: $cm$^{-2}$)
since usually several LB's are
occupied, which may overlap near $E_F$.
The energy spectrum has been investigated in the intermediate
region in the absence of collision broadening. The coupling of the LL's
by the external periodic potential
strongly reduces the original high symmetry of the
Hofstadter spectrum but retains a very complicated
subband structure \cite{Kuhn93:8225,Petschel93:239}.
All the above-mentioned theoretical investigations of the energy spectrum
of a 2DEG in a periodic potential and a homogeneous perpendicular
magnetic field have neglected the effects of the electron-electron
interaction. Here we study, within the Hartree
approximation (HA), these effects on
the electronic energy spectrum and the thermodynamic density of states.

To describe the electrons in the conduction band
in the presence of a lateral
superlattice at the AlGaAs-GaAs interface in a constant perpendicular
magnetic field $\vec B=B\hat z$ we employ a model of a
2DEG, with the three-dimensional charge density
given by $-e n_s(\vec{r})\delta(z)$, and $\vec{r}=(x,y)$. The square
superlattice is spanned by orthogonal lattice vectors
of length $L$, the primitive
translations of the Bravais lattice ${\cal B}$.
The corresponding reciprocal lattice ${\cal R}$ is spanned by
$\vec G=G_1\vec g_1 + G_2\vec g_2$, with $G_1, G_2 \in Z$,
$\vec g_1=2\pi{\hat x}/ {L}$, and $\vec g_2=2\pi{\hat y}/ {L}$.
The external periodic potential the electrons are moving in
is taken to be of the simple form
$V(\vec r)=V_0\{ \cos (g_1x) + \cos (g_2y)\} $.
The electron-electron interaction is included in the
HA leading to an effective single-electron Hamiltonian
$H=H_0+V_H(\vec r)+V(\vec r)$,
where $V_H(\vec r)$ is the effective potential in a medium with
a dielectric constant $\kappa$,
felt by each electron and caused by the total charge density of the
2DEG, $-en_s(\vec r)$, and the neutralizing background charge density
$+en_b=+e\langle n_s(\vec r)\rangle$.
The periodic external potential $V(\vec r)$ and the constant external
magnetic field imply that all physical quantities of the
noninteracting system are periodic with respect to translations of
$\vec R\in {\cal B}$.
The periodicity of the Hartree potential
follows from that of $n_s(\vec r)$.
In the following calculations we use the symmetric
basis functions constructed by
Ferrari \cite{Ferrari90:4598} and used by
Silberbauer \cite{Silberbauer92:7355}.
The set of Hartree equations has to be solved
iteratively together with the condition that the average electron
density $n_s=N_s/A$ is constant, which determines the chemical potential
$\mu$. $N_s$ is the number of electrons per unit cell
with area $A$ in ${\cal B}$.

The Coulomb interaction not only couples directly the subbands of a
particular LL, but also the subbands of different LL's.
If only the basis states in the same LL are taken into account,
the modulation strength is irrelevant since it can be factored out of the
Hamiltonian matrix.
In this situation, corresponding to the usual discussion of Hofstadter's
butterfly on the basis of Harper's equation, a highly symmetric energy
spectrum is expected. Of course, this restricted
model is only appropriate for describing a system with
very weak modulation (as compared with the inter-Landau-level
energy spacing $\hbar\omega_c$).
The bandwidth of the subbands of the first LB
is shown in Fig.\ \ref{Fig-BW50}, for the two cases of 3 or 6 magnetic
flux quanta through the unit lattice cell, $\Phi /\Phi_0 =pq=3,6$,
identifiable from the number of subbands appearing in each figure.
The period is short, $L=50\: $nm, the modulation strong,
$V_0=4.0\: $meV, and $T=1.0\: $K.
However only one LL is used in the calculation, so that the results apply to
the case of a weak superlattice
potential. The electron-hole symmetry is exact,
as is reflected in each case by the point symmetry of the
subband structure around
$\mu$ at half filling.
The screening is strongly dependent on $N_s$ and
the average filling factor. It is, therefore, strongest for $pq=6$ when
maximum six electrons can occupy the lowest Landau level
in the unit cell and when
the LB is half filled, $N_s=3$. The (dimensionlessly written)
thermodynamic density of states
$D_T=l^2\hbar\omega_c( {\partial n_s}/ {\partial\mu})_{TB}$, with
$n_s=\langle n_s(\vec r)\rangle$, is shown in
Fig.\ \ref{Fig-TDOS}. Clearly the strong screening in the $pq=6$
case obscures the fine structure in $D_T$ leading to only one
pronounced peak. The subbands in the other cases turn up as separate peaks
in $D_T$.
The Hofstadter butterfly should be recovered when the calculation is
performed for $N_s=0$ and only one LL, and
if the width and location of the subbands are plotted as functions of the
inverse of the number of magnetic flux quanta
through a unit cell of the lattice, i.e. of $\Phi_0/\Phi=1/(pq)$.
In addition, all energies have to be scaled according to:
$E\rightarrow (E-\hbar\omega_c/2)/V_0$. For interacting electrons
$N_s$, the average filling
factor of the LB turns out to be a new parameter controlling the
bandstructure. We therefore, present in Fig.\ \ref{Fig-TDOS} the subband
structure for
four values of the inverse flux $1/(pq)$ in the case of
$N_s=1.50$  together with
the location of $\mu$.
For low density, the subband structure is almost symmetric
around the energy zero, like in the noninteracting case, but the subbands
become quite
asymmetric for a higher density of electrons.
For the short-period superlattice
studied here, and for a low density of electrons, the essential gap structure
does survive  in the presence of interaction. The presentation in
Fig.\ \ref{Fig-TDOS}  corresponds to the
experimental procedure of keeping the density of electrons fixed
but changing the magnetic field.
Another way to investigate
the screening is to keep the average filling factor $\nu$ constant but
vary the magnetic field and, thus, also the average density
of the 2DEG. Fig.\ \ref{Fig-Hof} compares the subband structures
for $1/(pq)=1/2,1/3,1/4,1/6$ and $\nu =1/2$ with the complete
Hofstadter butterfly. Here the energies have been scaled with
the factor $(E-\hbar\omega_c/2)\exp\{ (\pi l/L)^2\} /V_0$ so that the results
can be directly
compared with earlier calculations of the Hofstadter
spectrum \cite{Pfannkuche92:12606}.
The energy spectrum in the interacting case
shows an overall reduction in dispersion or bandwidth due to the
strong screening that is most effective for large flux and large number of
available states. The bandwidths of the subbands for the interacting
case has been evaluated here on a discrete lattice
in the magnetic Brillouin zone without attempting an interpolation
between the lattice points; thus, the actual bandwidths can be
larger by a small percentage of the widths shown.

In an interacting 2DEG subject to a superlattice potential and a homogeneous
perpendicular magnetic field not only the magnetic flux through a
unit cell but also the density of electrons determines the
complicated splitting of the LL's into subbands.
We have shown that in the HA the essential
gap structure of the energy spectrum remains,
although the screening leads to a quenching of the Hofstadter butterfly
at small values of the inverse flux. The symmetry of the
butterfly is also lowered by
the coupling to higher LL's due to increased strength of
the periodic potential,
as has already been discussed by other
authors \cite{Kuhn93:8225,Petschel93:239}.
For periods around $L=200\: $nm (currently attainable in
experiments on superlattices) the 2DEG can effectively screen the
periodic potential even for a very low density $n_s$ when only one
or two Landau bands are partly occupied. Only at shorter
lattice constants ($L<100\: $nm) and thus much higher magnetic fields
we can, on the basis of our Hartree calculation, expect the subband
structure to be resolvable in experiments, when $n_s$
is maintained low enough. These predictions are made on
the basis of the calculated
energy spectrum and the structure observed in the thermodynamic density
of states $D_T$. The transport coefficients
do not depend on $D_T$ in any simple way,
and may be more sensitive to the subband structure of the energy
spectrum \cite{Pfannkuche92:12606}.
In order to keep the computational efforts in reasonable limits, we have
restricted our calculations to a few characteristic values of the magnetic
field. Our results indicate in which manner screening effects will change
the overall appearence of
the Hofstadter energy spectrum. Unfortunately, there is no
simple way to extrapolate from these special
values to arbitrary rational numbers of flux quanta per unit cell.
{}From the
experimental point of view these details of the energy spectrum seem not
to be accessible anyway. The challenge for the near future is to
resolve the most prominent gaps of the energy spectrum experimentally.

This research was supported in part by the Icelandic Natural Science
Foundation, the University of Iceland Research Fund,
and a NATO collaborative research Grant No. CRG 921204.
\newpage
\begin{figure}
\caption{The bandwidth of the minibands of the lowest Landau
         band as a function of $N_s$ for a Hartree interacting 2DEG and
         ($p=1$, $q=3$) (left), ($p=2$, $q=3$) (right).
         The chemical potential $\mu$ is indicated by a continuous curve.
         $L=50\: $nm, $T=1.0\: $K, $V_0=4.0\: $meV.
         One Landau level is included in the HA. $m^*=0.067m_0$,
         $\kappa =12.4$.
}
\label{Fig-BW50}
\end{figure}
\begin{figure}
\caption{The thermodynamic density of states (left)
         $D_T=l^2\hbar\omega_c(\partial n_s/\partial\mu)_{TB}$
         as a  function of $1/pq$ and $N_s$.
         The scaled bandwidth (right) of the subbands
         of the lowest Landau band,
         $(E-\hbar\omega_c/2)/V_0$, as function
         of $1/pq$ for $N_s=1.5$.
         The chemical potential $\mu$ is indicated by crosses.
         $L=50\: $nm, $T=1.0\: $K, $V_0=4.0\: $meV.
         One Landau level is included in the HA.
}
\label{Fig-TDOS}
\end{figure}
\begin{figure}
\caption{The scaled bandwidth $(E-\hbar\omega_c/2)\exp\{ (\pi l/L)^2\} /V_0$
         of the subbands of the lowest Landau band as function
         of $1/pq$ for the Hartree
         interacting 2DEG with $\nu=1/2$ (left), and for the
         noninteracting 2DEG (right), the Hofstadter butterfly.
         One Landau level is included in the calculations.
         In the left subfigure the 6 minibands for the case $pq=6$ can
         not all be resolved due to vanishing band gaps.
         The parameters used for the left subfigure are:
         $L=50\: $nm, $T=1.0\: $K, $V_0=4.0\: $meV.
}
\label{Fig-Hof}
\end{figure}
\end{document}